%% file: WLJFM.tex
\documentclass[referee]{jfm}
\usepackage{psfig,epsf,amsmath,amssymb,natbib}
\usepackage{graphicx,epsf,psfig}
\usepackage{bm}
\newcommand{\boldtau}{\mbox{\boldmath $\tau$}}

\newcommand{\boldalpha}{\mbox{\boldmath$\alpha$}}
\newcommand{\bolddelta}{\mbox{\boldmath$\delta$}}

\newcommand{\boldnabla}{\mbox{\boldmath$\nabla$}}

\newcommand{\Wi}{\textit{Wi}}
\newcommand{\Rey}{\textit{Re}}
\newcommand{\Ex}{\textit{Ex}}
\newcommand{\El}{\textit{El}}

\title[Viscoelastic nonlinear traveling waves]{Nonlinear traveling waves as a framework for understanding turbulent drag reduction}

\author{Wei Li, Li Xi and Michael D.~Graham}
\affiliation{%
Department of Chemical and Biological Engineering\\
 University of Wisconsin-Madison,
Madison, WI 53706-1691
}%
\date{\today}
\begin{document}
\maketitle
\input{abstract}


\section{Introduction}
\input{intro}
\section{Formulation}
\input{formulation}
\section{Results and discussion}
\input{results}

\input{conclusions}
\section*{Acknowledgments}
\input{acknowledge}
\bibliographystyle{jfm}
\bibliography{WLJFM}

\end{document}

%% file: abstract.tex
\begin{abstract}

    Nonlinear traveling waves that are precursors to laminar-turbulent
    transition and capture the main structures of the turbulent buffer
    layer have recently been found to exist in all the canonical
    parallel flow geometries.  We study the effect of polymer
    additives on these ``exact coherent states" (ECS), in the plane
    Poiseuille geometry.  Many key aspects of the turbulent drag
    reduction phenomenon are found, including: delay in transition to
    turbulence; drag reduction onset threshold; diameter and
    concentration effects.  Furthermore, examination of the ECS
    existence region leads to a distinct prediction, consistent with
    experiments, regarding the nature of the maximum drag reduction
    regime.  Specifically, at sufficiently high wall shear rates,
    viscoelasticity is found to completely suppress the normal (\emph{i.e.}
    streamwise-vortex-dominated) dynamics of the near wall region,
    indicating that the maximum drag reduction regime is dominated by
    a distinct class of flow structures.

\end{abstract}

%% file: intro.tex
The reduction of turbulent drag by polymer additives has received much
attention since it was first observed experimentally in 1940s~(See
reviews by \cite{lumley69,virk75,mccomb,graham04}).  For a given flow
rate, small polymer concentrations, on the order of ten parts per
million by weight, can reduce the pressure drop in pipe or channel
flow, for example, by 50$\%$ or greater.  After six decades of
research, the subject remains an active area of research, in part
because of applications but also because it lies at the intersection
of two complex and important fields, turbulence and polymer dynamics.
A better understanding of this phenomenon may in turn yield insights
into both the dynamics of drag-reducing fluids and of turbulent flows.
The goal of the present work is to address turbulent drag reduction in
the context of the dominant structures in the turbulent buffer layer,
an approach which turns out to touch on many key aspects of the drag
reduction phenomenon.

We focus here on pressure-driven channel flow with average wall shear
stress ${\tau}_{w}$, of a fluid with dynamic viscosity $\eta_{s}$ (in
the absence of polymer), density $\rho$ and kinematic viscosity
$\nu={\eta}_{s}/\rho$.  The average streamwise velocity
$U_{\text{avg}}$ and half-channel height $l$ define outer scales for
the flow.  Inner scales are the friction velocity
$u_{\tau}=\sqrt{\tau_{w}/\rho}$ and the near-wall length scale $l_{w}=
\nu/u_{\tau}$.  As usual, quantities expressed in terms of these
so-called ``wall units" are denoted with a superscript ${}^{+}$.  The
friction Reynolds number $\Rey_{\tau}=u_{\tau} l /\nu$ is simply the
half channel height expressed in wall units.  The Weissenberg number
is denoted $\Wi=\lambda {\dot{\gamma}}_w=\lambda u_{\tau}^{2}/\nu$,
where $\lambda$ is polymer relaxation time and ${\dot{\gamma}}_w$ is
the average wall shear rate.  Experimental results for a given fluid
and flow geometry lie on curves of constant elasticity parameter
$\El=2\lambda\nu/l^2$.

\begin{figure}
\centerline{\includegraphics[width=2.75in]{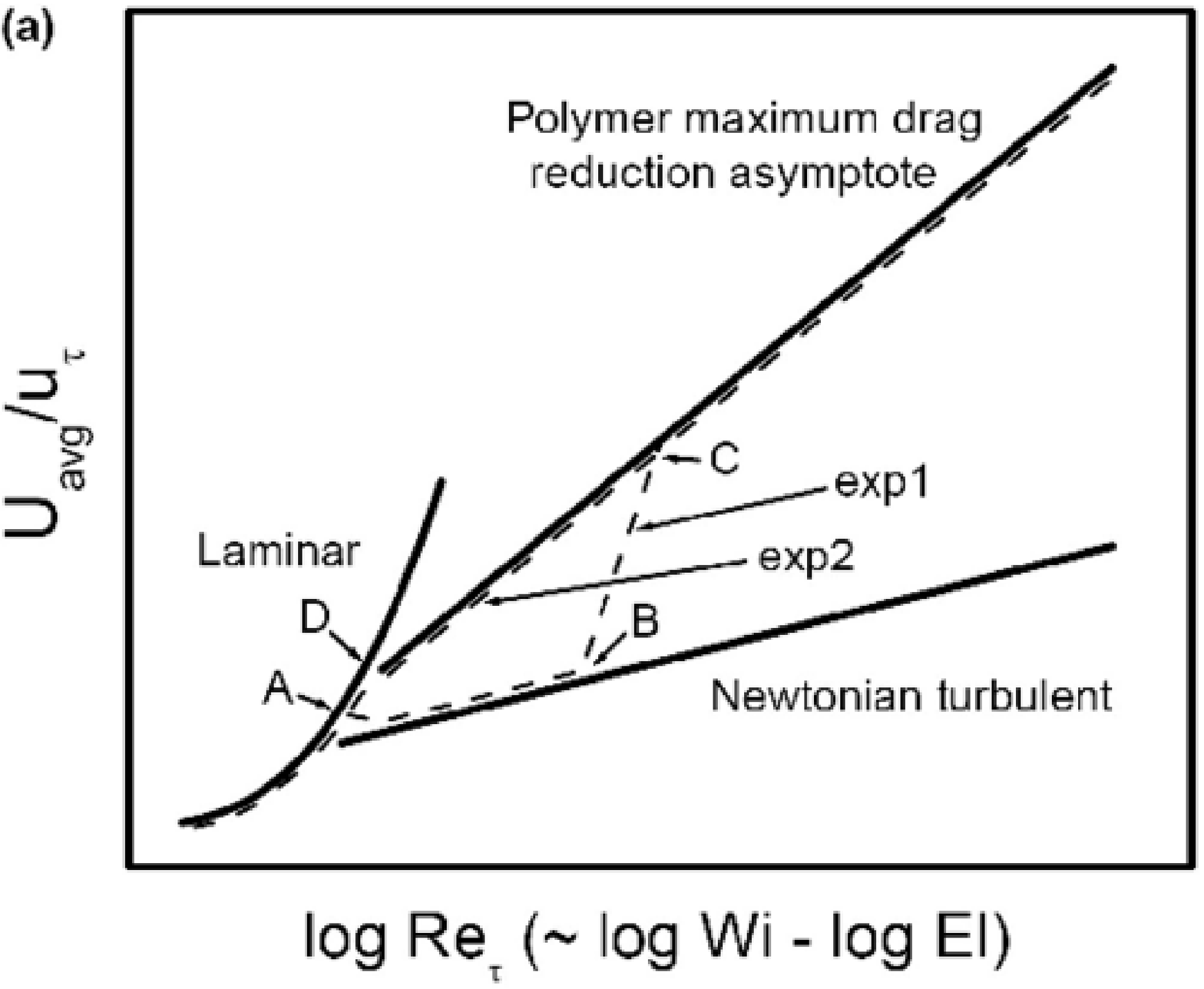}\includegraphics[width=2.75in]{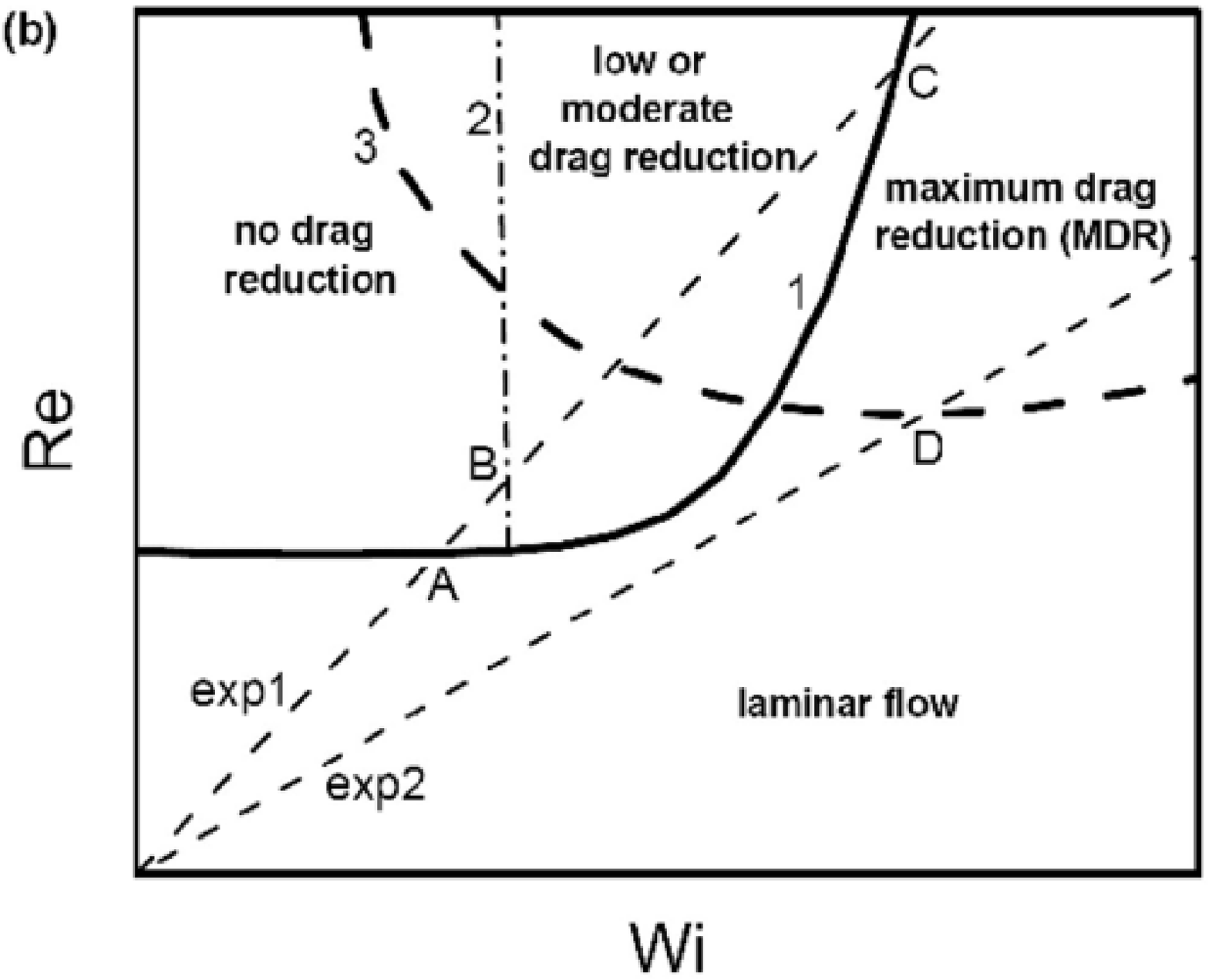}}
\caption{(a) Schematic Prandtl-von Karman plot.  The dashed lines
represent the experimental paths by which specific polymer systems
of different molecular weights, concentrations, polymer-solvent
pairs, etc., approach the maximum drag reduction asymptote.  (b)
Schematic of polymer induced turbulent drag reduction based on
existence regions for nonlinear coherent states.} \label{fig:PVK}
\end{figure}

In channel or pipe flow, drag reduction results are often
represented on a Prandtl-von Karman plot,
$U_{\text{avg}}/u_{\tau}$ \emph{vs.} $\log \Rey_{\tau}$ ($\sim
\log \Wi-\log \El$), shown schematically in Fig.~\ref{fig:PVK}a. Point A
corresponds to transition to turbulence, which in Newtonian flow occurs
at $\Rey_{\tau}\approx 45$ \citep{carlson82}.  One typical
experimental path for a given polymer solution and channel size is shown by the curve labeled
``exp1''.  Along this path, once $\Rey_{\tau}$
exceeds a critical value (point $B$), the slope of the data
increases from the Newtonian value, indicating onset of drag
reduction.
As $\Rey_{\tau}$ increases, data eventually approaches a new curve at
point $C$.  This curve, the so-called Maximum Drag Reduction (MDR)
asymptote, is insensitive to polymer concentration, molecular weight
or polymer species -- all results collapse onto it at large
$\Rey_{\tau}$; it is a universal feature of drag reduction by
polymers.  For large channels or low polymer concentrations, the value
of $\Rey_{\tau}$ at the onset of drag reduction is independent of
polymer concentration and corresponds to a critical Weissenberg
number.  For small channels or large concentrations, however, diameter
and concentration effects have been observed
experimentally~\citep{virk75}: specifically, there exists a critical
pipe diameter below which, or a critical polymer concentration above
which the flow behavior directly transits from laminar flow to the
maximum drag reduction curve as $\Rey_{\tau}$ increases.  An
experimental path showing this effect is labeled ``exp2''; transition
from laminar flow to MDR occurs at point $D$.

Studies of drag-reducing fluids indicate that near the onset
of  drag reduction, the effects of the polymer
are confined primarily to the buffer layer region of the
flow~\citep{virk75,tiederman72}.
Experimental observations and DNS studies show that the dominant
structures in the buffer layer are pairs of counter-rotating,
streamwise-aligned vortices~\citep{robinson91,jeong97}. These
vortices pull slower moving fluids away from the wall, forming
low-speed, streamwise velocity streaks. In drag-reducing flows,
these structures are modified by polymers: the buffer region
thickens~\citep{virk75}, the coherent structures in this region
shift to larger scales~\citep{tiederman72,beris97,nieuwstadt97},
and the bursting rate decreases~\citep{tiederman72}. Recent
experimental results~\citep{Warholic99,Warholic01} reveal that in
the maximum drag reduction region the ejections from the wall are
eliminated and the near-wall vortices that sustain turbulence in a
Newtonian fluid are completely destroyed. Low-speed streamwise
velocity streaks are essentially absent. These observations
suggest that the coherent structures in buffer layer region are
crucial in addressing rheological drag reduction in wall-bounded
turbulent flows.

 A recent advance in the understanding of these important near-wall
 structures has come with the recognition that, in all the canonical
 parallel geometries (plane Couette, plane Poiseuille, pipe) the
 Navier-Stokes equations support nonlinear traveling wave states, the
 family of so-called ``exact coherent states" or ECS
 \citep{nagata86,busse97,waleffe98,waleffe01,waleffe03,eckhardt03,WedinKerswell04}.
 Jim\'enez and coworkers \citep{pinelli99,jimenez01} have found related states in
 spatially filtered direct numerical simulations (DNS), showing the
 autonomous nature of the near-wall behavior.  The flow structure of
 these states is a mean shear and a pair of staggered
 streamwise-aligned counter-rotating vortices, as is found in the
 turbulent buffer layer.  In the plane Poiseuille geometry, ECS come
 into existence at $\Rey_{\tau}$ of 44.2 \citep{waleffe03}, very close
 to the experimentally observed $\Rey_{\tau}$ of $\sim 45$ for the
 transition to turbulence\citep{carlson82}.  The spanwise wavelength
 $L_z^+ = 105.5$ of the ECS at onset closely matches the streak
 spacing of $\sim 100$ wall-units widely observed in experiments over
 a large range of Reynolds numbers~\citep{robinson91}.  Direct
 numerical simulations of turbulence in ``minimal channel
 flow'', \emph{i.e.}, flow in the smallest computational domain that
 reproduces the velocity field statistics of near-wall turbulence,
 give a range for the streamwise length $L_{x}^{+}$ of $250-350$,
 compared to $L_x^+ = 273.7$ for the ECS, and a spanwise length that
 is again approximately $100$ wall units~\citep{jimenez91}.
It should be pointed out that this minimum channel contains a
single wavelength of a wavy streak and a pair of quasi-streamwise
vortices, which is the same structure seen in the ECS. A conditional
sampling study of coherent structures in a larger scale
DNS~\citep{jeong97} indicates that the dominant structures near the
wall in turbulent channel flow are counter-rotating,
streamwise-aligned vortices with a streamwise length $L_{x}^{+} \sim
250$, a spanwise length $L_{z}^{+} \sim 100$ and a wall-normal size of
$y^{+} \sim 50$, which agrees with the scales of the ECS at onset.
The ECS also capture the location of the peak, at $y^{+} \approx 12$, in
the production of turbulent kinetic energy for wall-bounded
turbulence~\citep{Kim87,LiGraham05}.  In short, the ECS are precursors
to turbulence and their structure and length scales closely match
experimentally observed near-wall behavior.

Because the first effects of polymer arise in the buffer region,
whose structure the ECS evidently capture, these flows provide a
natural starting point for understanding drag reduction. In prior
work, we have studied the initial effects of viscoelasticity on
ECS in the plane Couette and plane Poiseuille
geometries~\citep{stoneprl,stonebd,StoneGraham04,LiGraham05}.  The
primary effect was found to be the weakening of the streamwise
vortices, as well as changes in the statistics of the velocity
fluctuations that are consistent with experimental observations at
low levels of drag reduction.  The present work takes a broader
view, examining the region of parameter space ($\Rey$, $\Wi$) in
which ECS exist and its connection to experimental observations.


%% file: formulation.tex
We consider pressure-driven flow with no-slip boundary conditions;
$v_x$, $v_y$, and $v_z$ are streamwise, wall-normal, and spanwise
components of the velocity, ${\bf v}$, respectively. Reflection
symmetry is imposed at the channel centerline.
The laminar centerline velocity, $U$, and the half-channel height,
$l$, are used to scale velocity and position, respectively.  The
average wall shear rate $\dot{\gamma}_{w}$ is given by $2U/l$.  Time,
$t$, is scaled with $l/U$, and pressure, $p$, with ${\rho}U^2$.  The
stress due to the polymer, $\boldtau_p$, is nondimensionalized with
the polymer elastic modulus, $G = \eta_p/{\lambda}$, where $\eta_p$ is
the polymer contribution to the zero-shear rate viscosity.  The
momentum balance and the equation of continuity are
\begin{eqnarray}\label{NSE}
\frac{D {\bf v}}{D t} = - \boldnabla p + \beta \frac{1}{\Rey}
{\nabla^2}{\bf v}
+ (1-\beta) \frac{2}{\Rey {\Wi}}(\boldnabla \cdot \boldtau_p),\\
\label{continuity} \boldnabla \cdot {\bf v} = 0.
\end{eqnarray}
Here $\beta = \eta_s / (\eta_s + \eta_p)$ is the fraction of the total
zero-shear viscosity that is due to the solvent,
$\Rey=\frac{{\rho}{U}{l}}{\eta_{s}+\eta_{p}}$ and
$\Rey_{\tau}=\sqrt{2\Rey}$.

The polymer stress is computed with the widely-used FENE-P
constitutive model \citep{dpl2}:
\begin{equation} \label{alphaequ}
\frac{\boldalpha}{1-\frac{tr \boldalpha}{b}} + \frac{\Wi}{2}
\left( \frac{D {\boldalpha}}{D t} - {\boldalpha} \cdot \boldnabla
{\bf v} - {\boldnabla {\bf v}}^T \cdot {\boldalpha} \right) =
\frac{b \bolddelta}{b+2},
\end{equation}
where $\boldalpha$ is a non-dimensional conformation tensor and $b$
is proportional to the maximum extension of the dumbbell --- $tr
{\boldalpha}$ cannot exceed $b$.  The polymer contribution to the
stress is given by:
\begin{equation}
    \label{taupequ}
    \boldtau_p = \frac{b+5}{b}\left(\frac{\boldalpha} {1-\frac{tr
    \boldalpha}{b}} - \left( 1 - \frac{2}{b+2} \right) \bolddelta
    \right).
\end{equation}

The extensibility parameter $\Ex=\frac{2b(1- \beta)}{3 \beta}$
measures the relative magnitude of the polymer and solvent
contributions to the steady state extensional stress in uniaxial
extension at high extension
rate.  
We consider the situation $1-\beta\ll 1$, in which case shear-thinning
is negligible, as the polymer only contributes a very small amount to
the total shear viscosity of the solution.  In this situation,
significant effects of the polymer on the flow are expected only when
$\Ex\gg 1$.  Finally, recall that experimental
results for a given fluid and flow geometry lie on curves of constant
elasticity parameter $\El=2\lambda\nu/l^2=\Wi/\Rey$.

The conservation and constitutive equations are solved through a
Picard iteration in a traveling reference frame -- the wave speed is
part of the solution.  A Newtonian ECS, as computed in
~\cite{waleffe98}, is first used to calculate the polymer stress
tensor, ${\boldtau}_p$, by inserting the velocity field in the
evolution equation for ${\boldalpha}$ and integrating for a short
length of time, usually one time unit ($l/U$). For this
${\boldtau}_p$, a steady state of the momentum and continuity
equations is found by Newton iteration. The resulting velocity
field, ${\bf v}$, is used to compute the new ${\boldtau}_p$, and
the process is repeated until the velocity and polymer field
converge to a steady state.

The momentum and continuity equations are discretized using a
Fourier-Chebyshev formulation with typically a
$9{\times}17{\times}9$ grid. The conformation tensor,
${\boldalpha}$, is discretized with a third-order, compact upwind
difference scheme~\citep{lele92,choi01} in the $x$ and $z$
directions and 
Chebyshev collocation 
in the $y$ direction.  In this as in most previous computational
studies of polymers in turbulent flows, we have found it necessary to
add an artificial stress diffusion term
$\frac{1}{ScRe}{\nabla}^2{\boldalpha}$, to the right-hand side of Eq.
(\ref{alphaequ}) to achieve numerical stability.  The Schmidt number,
$Sc$, which is the ratio of the momentum diffusivity to stress
diffusivity, is set to value of 1.0.  This value of $Sc$, though
artificially small, is greater or of the same order of magnitude as
that used in many DNS studies
~\citep{beris97,nieuwstadt03,beris95,baron02}.  In the range of $Sc$
where solutions can be obtained, the bifurcation diagrams shown below
are insensitive to its value.  The stress diffusion term is integrated
implicitly by the Crank-Nicholson method with the other terms of the
equation integrated using the Adams-Bashforth method.  This equation
is solved on a finer mesh than the momentum, continuity pair, typically
$48{\times}49{\times}48$.


%% file: results.tex
\begin{figure}
\centerline{\includegraphics[width=2.75in]{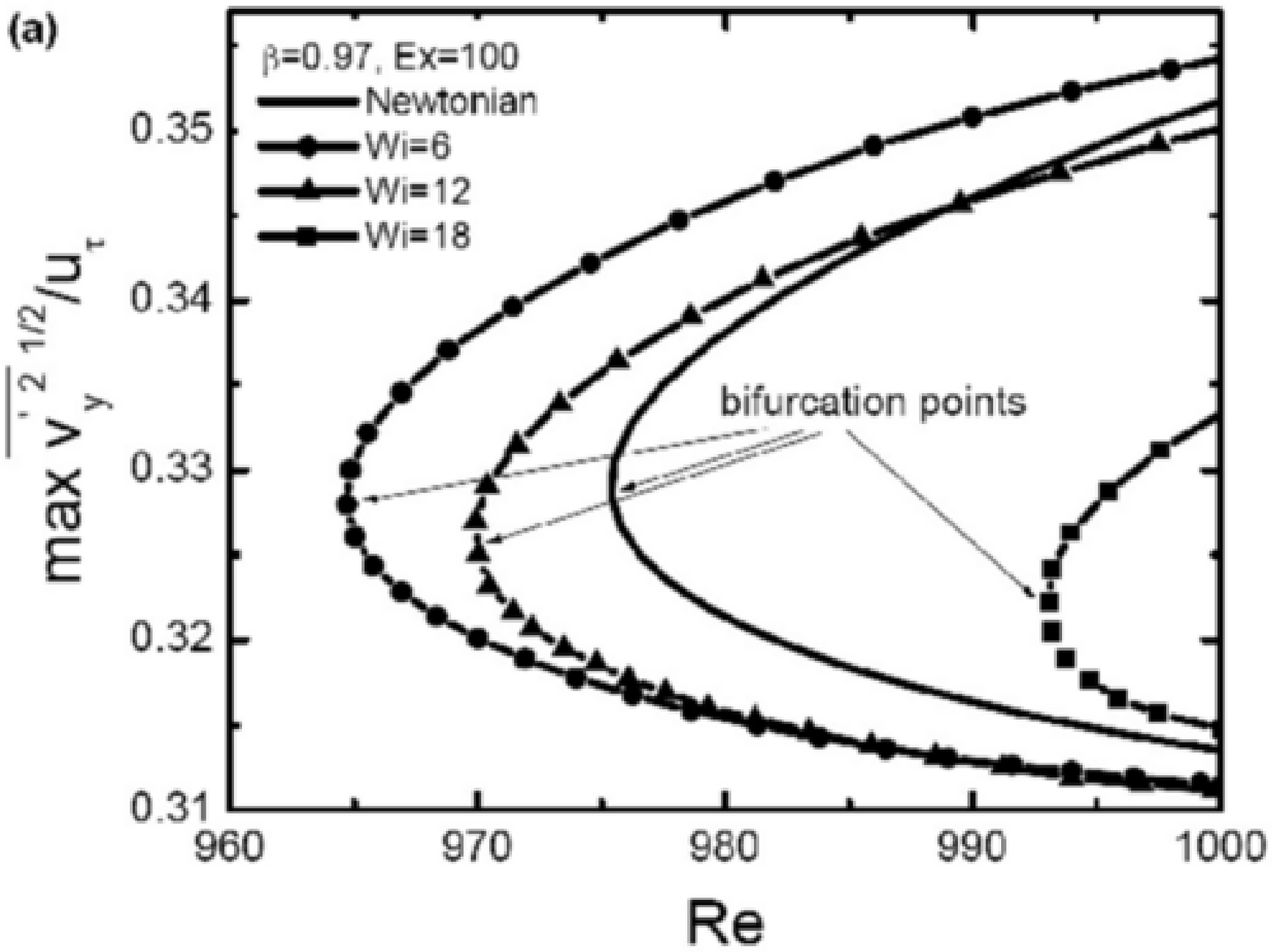}\includegraphics[width=2.75in]{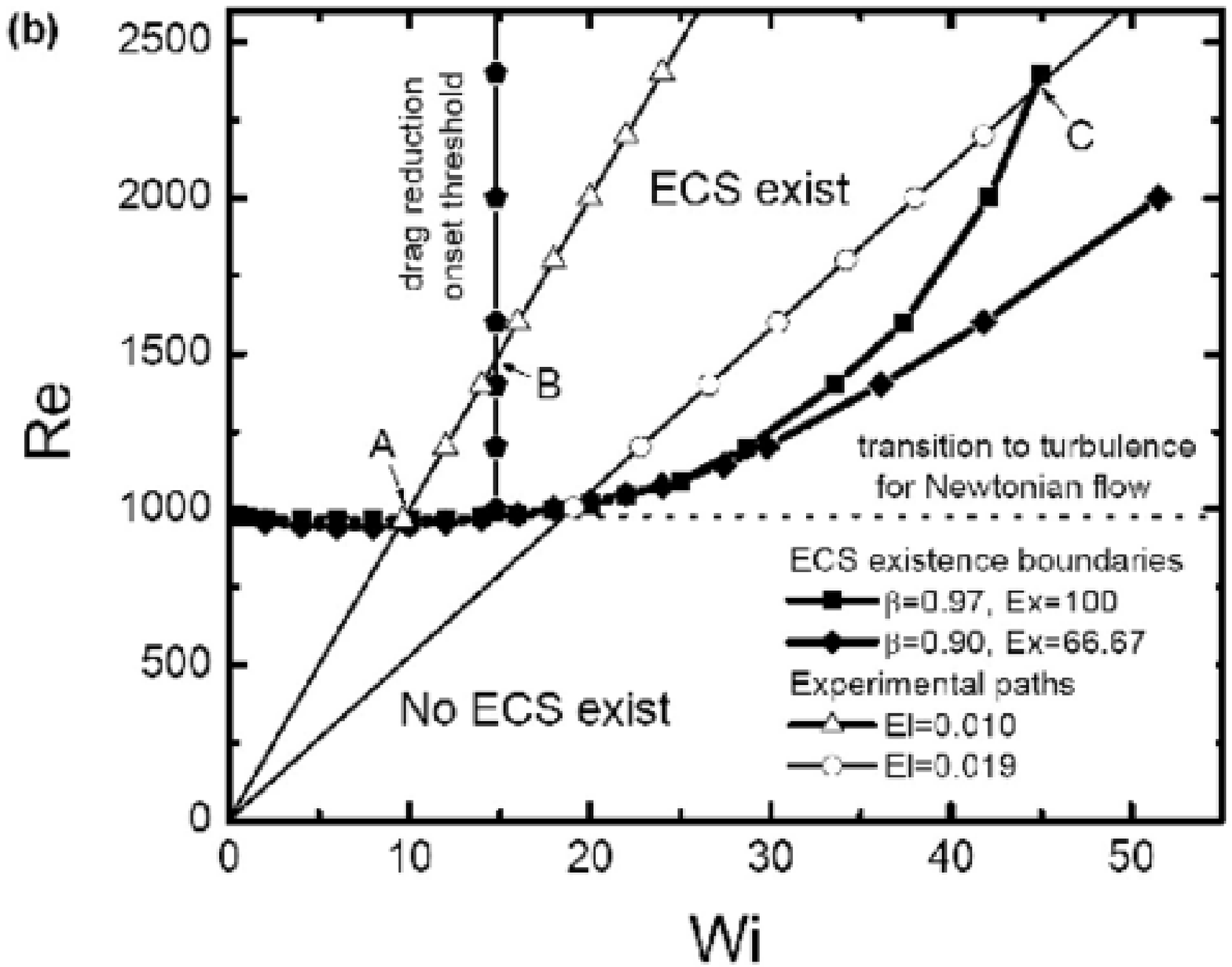}}
\caption{(a) Bifurcation diagram for Newtonian and viscoelastic ECS.
(b) Existence boundaries and drag reduction regimes for viscoelastic
ECS. For all results, $L_x=2\pi/1.0148$ and $L_z=2\pi/2.633$.}
\label{fig:bifur_exis}
\end{figure}

We study the Newtonian and viscoelastic ECS at fixed streamwise and
spanwise lengths: $L_x=2\pi/1.0148$ and $L_z=2\pi/2.633$ (\emph{i.e.},
$L_{x}^{+} = 273.7$ and $L_{z}^{+} = 105.5$ at $\Rey_{\tau}=44.2$).
This wavelength pair is where ECS first come into existence in the
Newtonian case.  The trivial base state in this geometry (laminar
Poiseuille flow) exists at all $\Rey$.  At $\Rey \approx 977$
($\Rey_{\tau}=44.2$) for the Newtonian flow, two new solutions appear
via a saddle-node bifurcation as shown in Fig.  \ref{fig:bifur_exis}a.  These are the ECS. These solutions are plotted using the maximum
in the root mean square wall-normal velocity fluctuations for the
solution, $\overline{v_y ^{'2}}^{1/2}$.  (Hereafter, an overbar
indicates that the variable is averaged over the streamwise and
spanwise directions.)  The solutions with higher maximum wall-normal
velocity at a given $\Rey$ are called ``high drag'' solutions due to
their lower mean velocity at the centerline of the channel compared to
the ``low drag'' solutions.  All results in this paper are for the ``high
drag'' states.  Although both solutions are unstable,
their status as precursors to transition and their structural
similarity to buffer layer turbulence suggest that they are saddle
points that underlie in part the strange attractor of turbulent flow.


Fig.~\ref{fig:bifur_exis}a indicates that the addition of polymer
changes the Reynolds number ${\Rey}_{\text{min}}$ at which the ECS
come into existence.  Curves of ECS existence boundaries
$\Rey_{\text{min}}$ vs.~$\Wi$ are given for two parameter sets
by the thick solid curves on Fig.  \ref{fig:bifur_exis}b.  These
separate the region where the ECS can exist (above the curves) from
the region where no ECS exist, for the given value of $\Ex$.  While at
low $\Wi$, there is a slight decrease in $\Rey_{\text{min}}$ from the
Newtonian value, once$\Wi$ exceeds 45, $\Rey_{\text{min}}$ for
$Ex=100$ is more than doubled.  This dramatic increase in
$\Rey_{\text{min}}$ after onset is consistent with the experimental
observation that the transition to turbulence in a polymer solution is
delayed to higher $\Rey$ than in the Newtonian
case~\citep{GilesPettit67,McEligot70,smith99}.  We will refer to the
$\Wi$ above which $\Rey_{\text{min}}$ for the viscoelastic ECS is
greater than for the Newtonian ECS as the onset Weissenberg number
$\Wi_{\text{onset}}$ for drag reduction.  Fig.  \ref{fig:bifur_exis}b shows that $\Wi_{\text{onset}} \approx 15$, a value which is
insensitive to polymer extensibility ($\Ex$) or concentration
($\beta$).  Furthermore, in simulations at constant $\Rey$, it is
found that the value where the centerline mean velocity
$U_{\text{max}}$ first exceeds the Newtonian value -- the nearly
vertical set of points in Fig.  \ref{fig:bifur_exis}b -- is also
located at $Wi \approx 15$ in the parameter regime that has been
examined here.  This onset value is high by about a factor of two
compared to values predicted by two recent viscoelastic DNS
studies~\citep{beris03,joseph03}, but in those studies $\El$ was
significantly smaller, and the onset Reynolds number correspondingly 
larger, than the values considered here.

\begin{figure}
\centerline{\includegraphics[width=2.5in]{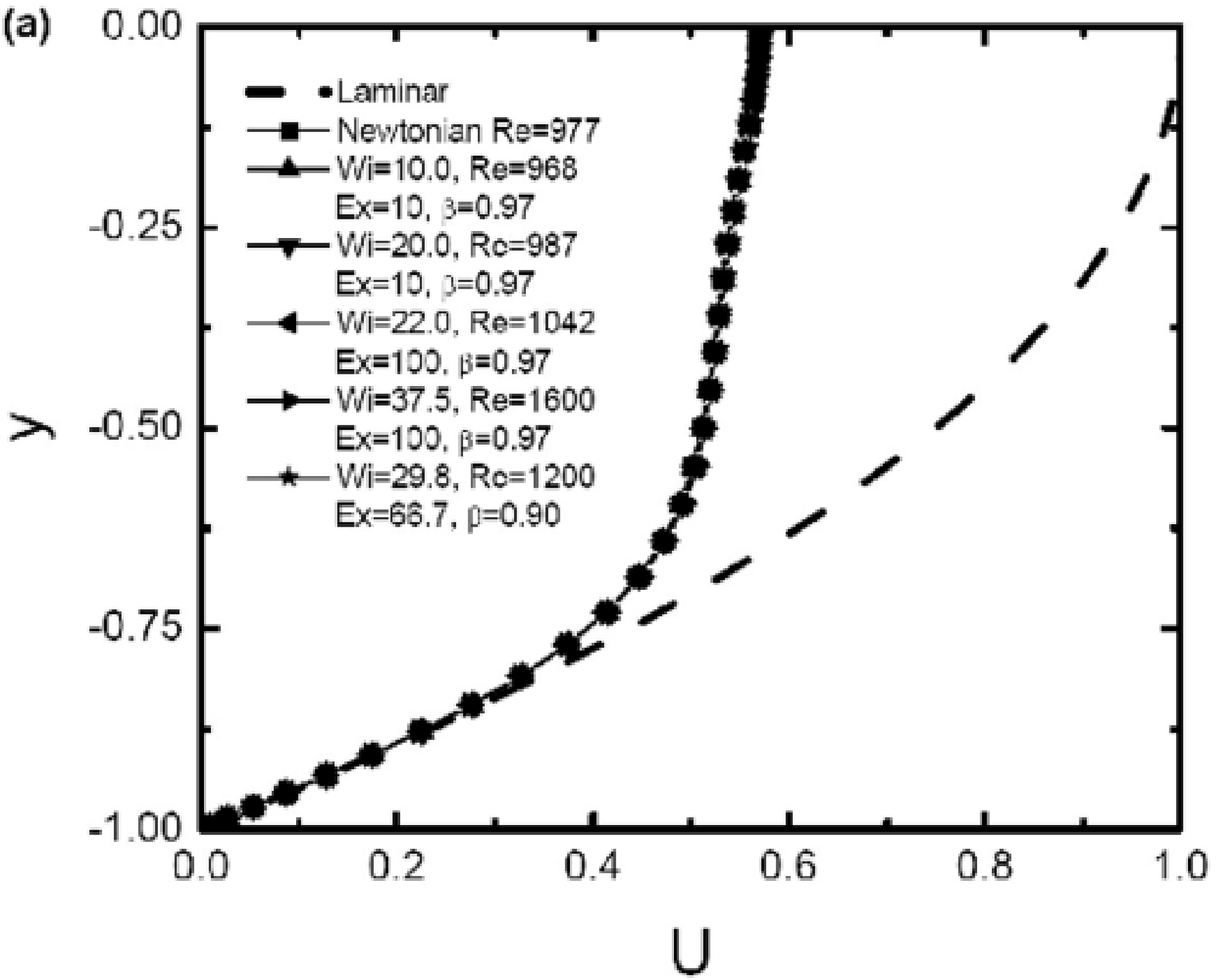}\includegraphics[width=2.5in]{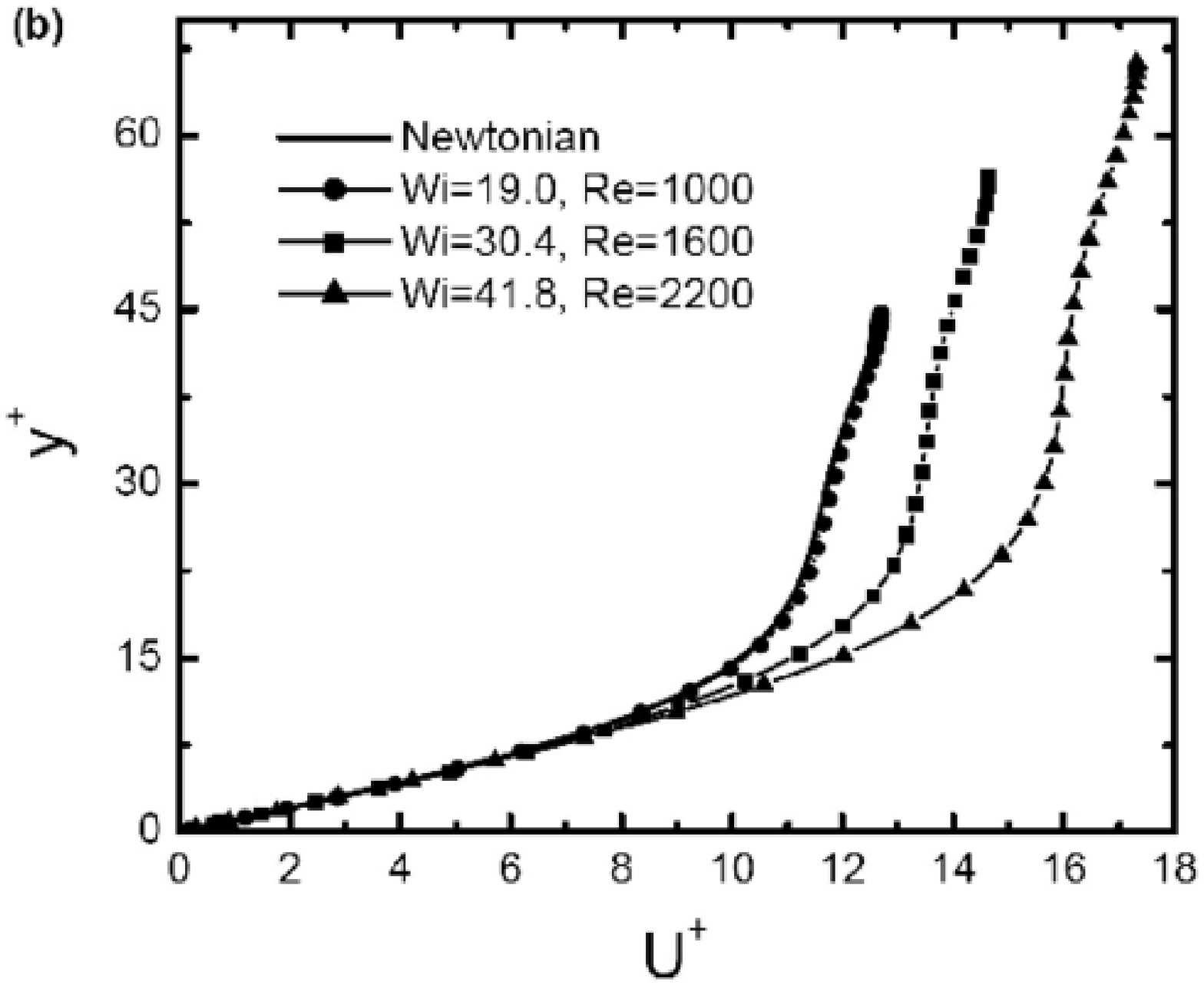}}
\caption{Mean streamwise velocity for Newtonian and viscoelastic
ECS on the ECS existence boundary.} \label{fig:Mean}
\end{figure}

Figure \ref{fig:Mean}a shows mean velocity profiles at six
different sets of parameter values, each
corresponding to a point on the existence boundary for the ECS (\emph{i.e.}
a
bifurcation point).  Remarkably, they all fall on the same
curve, when plotted in outer units.  Therefore, at least for the
values of $\Rey$ and $\Wi$ that are currently accessible in our
simulations, we observe that mean velocity profiles at onset of
the ECS have a universal form.

\begin{figure}
\centerline{\includegraphics[width=2.75in]{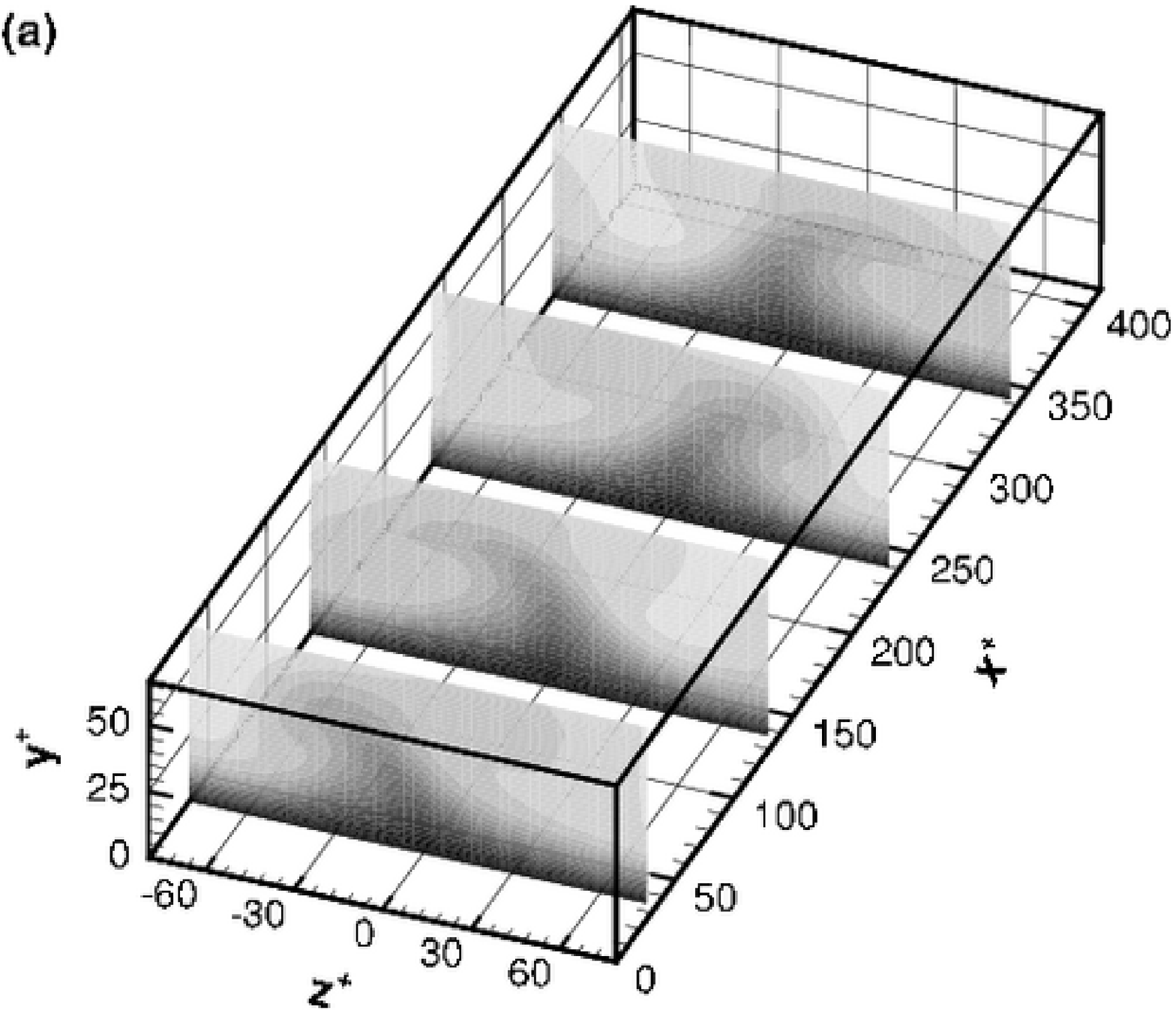}\includegraphics[width=2.75in]{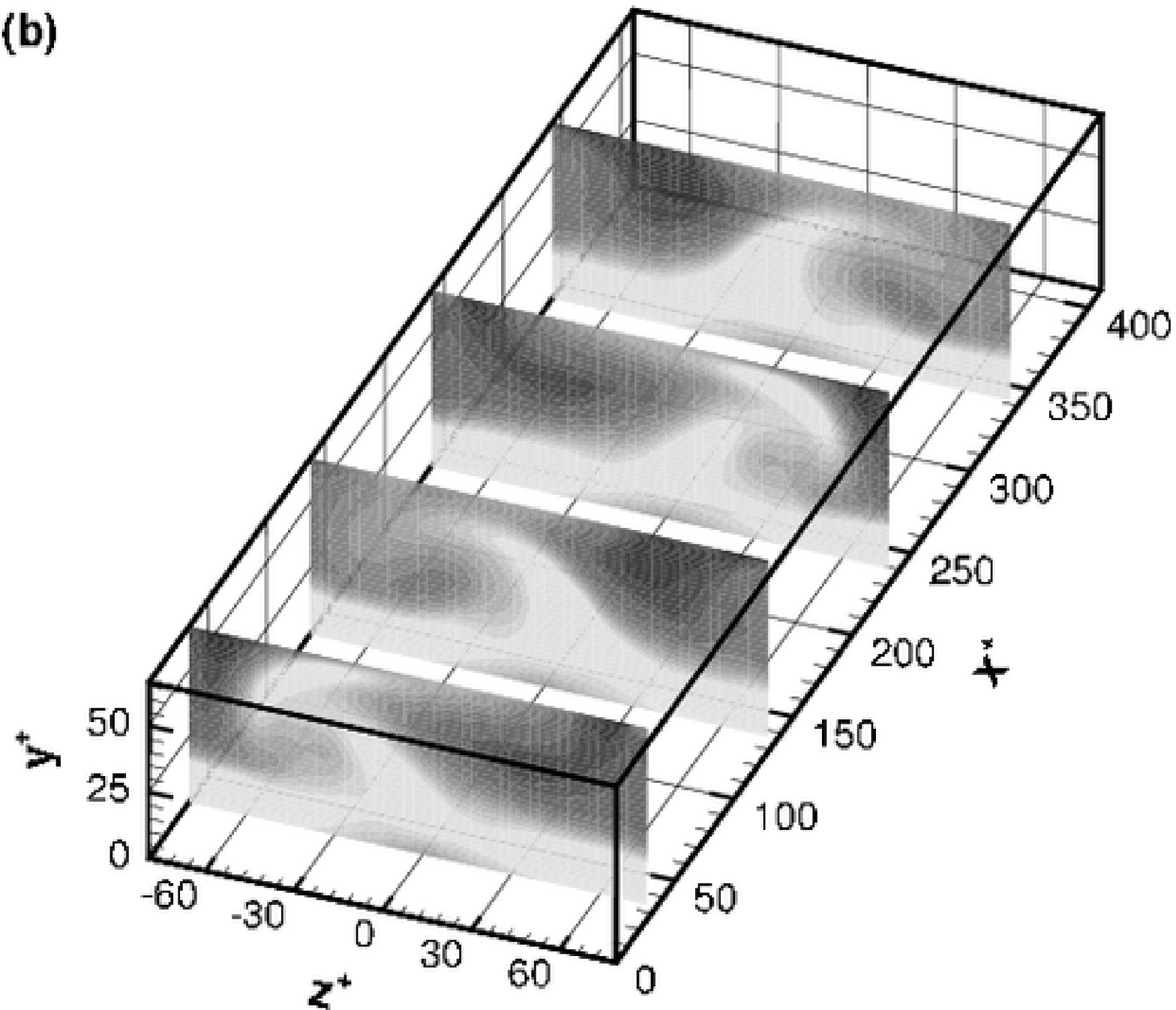}}
\caption{(a) Streamwise velocity for an exact coherent state at
$\Rey=2200 (\Rey_{\tau}=66.3)$, $\Wi=41.8$, $Ex=100$, $\beta=0.97$. 
Range: 0 (black) -- 0.58 (white).
(b) Trace of the polymer stress for the same state.  Range: 0
(black) --1800 (white).} \label{fig:velstress}
\end{figure}

We now turn to the study of the evolution of the ECS along some
experimental paths, lines of constant $El$.  Two such paths, the thin
solid lines with hollow symbols, are shown in
Fig.~\ref{fig:bifur_exis}b.  Consider first the case $\El=0.010$;
as $\Rey$ and $\Wi$ increase, the path intersects the ECS existence
boundary at point $A$ and the drag reduction onset threshold curve at
point $B$, where the transition to turbulence and the onset of drag
reduction occur, respectively.  Turning to the case $\El=0.019$, mean
velocity profiles expressed in wall units are shown for various values
of $\Rey$ in Fig.
\ref{fig:Mean}b.  For this parameter set, drag reduction is
observed immediately upon onset of the ECS. 
Along with drag reduction, enhanced streamwise velocity fluctuations
and the reduced wall-normal and spanwise velocity fluctuations are
found, consistent with experimental observations and DNS results at
low to moderate degrees of drag reduction \citep{virk75,beris97}.  The
effect of viscoelasticity can also be observed in the reduced Reynolds
shear stress and ultimately can be traced to the suppression of the
streamwise vortices by the viscoelasticity
\citep{stoneprl,StoneGraham04,LiGraham05}.  Figure \ref{fig:velstress}
shows fields of $v_{x}$ and $\text{tr} \boldtau_p$ on the $\El=0.019$ path
at $\Wi=41.8, \Rey=2200$ (the point just left of the label ``C'' on
Figure~\ref{fig:bifur_exis}b).  The region of high polymer stress clearly ``wraps 
around'' the streamwise vortices, and the corresponding polymer force
($\sim\boldnabla \cdot \boldtau_p$) is in direct opposition to the
vortex motions.

Continuing upward in $\Rey$ and $\Wi$ at $El=0.019$, the path
re-intersects the ECS existence boundary, at point $C$ in
Fig.~\ref{fig:bifur_exis}b.  (We suspect that this will also happen in
the $\El=0.010$ case, but at higher $\Rey$ and $\Wi$ than are
accessible with our current computational approach.)  Above this point
the flow can no longer sustain these ECS; viscoelasticity completely
suppresses the near-wall vortical structures.  This result is
consistent with experimental observations that, in the MDR regime, the
eruptions of low-momentum fluid from the wall are eliminated and the
near-wall streamwise vortices are completely
destroyed~\citep{Warholic99,Warholic01}.  Experimental results also
show that in the MDR regime, the Reynolds shear stress is much smaller
than the Newtonian value~\citep{Warholic99,Warholic01,HulsenMDR01},
and streamwise velocity fluctuations decrease to levels close to or
below the Newtonian value~\citep{Warholic99}.  All these observations
suggest that the turbulent production and dissipation take place by a
different mechanism in the MDR regime than at lower degrees of drag
reduction.  Although our study does not reveal this mechanism
directly, it does suggest that the disappearance of ECS is related to
the MDR regime.  This result encourages us to take a broader view,
examining the region of parameter space ($\Rey$, $\Wi$) in which ECS
exist and its connection to experimental observations.

Fig.~\ref{fig:PVK}b is a schematic based on the results shown in
Fig.~\ref{fig:bifur_exis}b.  Line 1 represents the ECS existence
boundary at constant $\Ex$.  Line 2 represents the drag reduction
onset threshold, which separates the ECS existence region into
``turbulence without drag reduction" and ``turbulence with low or
moderate drag reduction" regions.  Line ``exp1'' represents an
experimental path at constant $El$, which passes through the ECS
existence region.  In this case, as $\Rey$ (and $\Wi$) increases, this
path intersects with the ECS existence boundary at point $A$ and drag
reduction onset threshold at point $B$, where the transition to
turbulence and the drag reduction onset occur, respectively.  Note the
correspondence with points A and B on the schematic Prandtl-von Karman
plot, Fig.~\ref{fig:PVK}a, as well as on Fig.~\ref{fig:bifur_exis}b.
As $\Rey$ and $\Wi$ continue to increase along this path, the system
will eventually exit the ECS existence region at point $C$, where the
flow can no longer sustain these ECS. Experimental results show that
in the MDR regime, near-wall streamwise vortical structures are
essentially absent.  Our results together with this experimental
observation suggest that the loss of ECS may be somehow related to the
approach of the MDR regime, in which other types of coherent traveling
wave states (temporally intermittent structures, hairpins,
Tollmien-Schlichting waves, intrinsically elastic structures,\ldots) may be unmasked and become
dominant.  This possibility is represented by line 3 in
Fig.~\ref{fig:PVK}b, a hypothetical existence boundary for a distinct
class of flow structures that exists at high $\Wi$.  In this scenario,
the crossing of path exp1 across point C represents the transition to
the maximum drag reduction regime.
This scenario, incorporating transition to turbulence, onset of drag
reduction and approach of the MDR regime is consistent with the
behavior on experimental path exp1 shown in Fig.  \ref{fig:PVK}a.

Now consider the experimental path ``exp2'' on Fig.~\ref{fig:PVK}b.
This path corresponds to a value of $\El$ that does not intersect with
ECS existence region at all.  For the conditions $\beta=0.97, \Ex=100$
shown on Fig.~\ref{fig:bifur_exis}b, this situation arises if
$\El\gtrsim 0.024$.  The scenario on Fig.~\ref{fig:PVK}b would predict
in this case that, with the increase of $\Rey$ and $\Wi$, the flow
behavior directly transits from laminar to MDR at point $D$.
As $\El$ is inversely proportional to $l^{2}$ (or $R^{2}$ in pipe
flow) this prediction is consistent with experiments in small diameter
pipes -- the ``diameter effect"~\citep{virk75}, as exemplified by
experimental path exp2 in Fig.  \ref{fig:PVK}a.  The ``concentration
effect" can also be captured by this scenario, as we now describe.
The quantity $S=1-\beta$ is proportional to polymer concentration in
dilute solution.  Using $S$, the parameters $\Ex$ and $\El$ can be
written as $\Ex=\frac{2}{3}bS/(1-S)$ and $\El=2\lambda
\eta_{s}/\rho l^{2}(1-S)$.  Thus while $\El$ is virtually unchanged by
a change in $S$, $\Ex$ is proportional to it.  An increase in $\Ex$
compresses the ECS existence boundary leftward and eventually a
given experimental path can no longer intersect the ECS existence region, 
resulting again in flow behavior that directly transits from
laminar to MDR.

Finally, we observe that the existence boundaries can be interpreted in
terms of length scales.  Recall that the half-height of the channel,
expressed in wall units, is simply $\Rey_{\tau}=\sqrt{2 \Rey}$.  Thus the
existence boundary corresponds to the minimum half-channel height in which 
an ECS can exist, as a function of $\Wi$.  Points where a line of constant 
$\El$ intersects the existence boundary are points where the channel height 
and the minimum height for the existence of an ECS coincide.

\section{Conclusions}
Many observations of drag reduction in dilute polymer solutions are
mirrored by the effect of viscoelasticity on the channel flow ECS
discovered by Waleffe \citep{waleffe01,waleffe03}.  The transition behaviors
from laminar to turbulent flow, from no drag reduction to drag
reduction, and from moderate drag reduction to MDR can be connected to
the birth, evolution and death of these ECS, respectively.  Our
results and the scenario that we infer from them yield explicit
predictions, testable by DNS, with regard to all these phenomena.

%% file: acknowledge.tex
The authors are indebted to Fabian Waleffe for many illuminating
discussions and for sharing his code for computation of the
Newtonian exact coherent states.  
This work was supported by the
National Science Foundation, grant CTS-0328325, and the Petroleum
Research Fund, administered by the American Chemical Society.